\documentclass{article}
\usepackage{mathtools}
\usepackage{algorithm}
\usepackage{tabularx} 
\usepackage{array}
\usepackage{longtable}
\usepackage{balance}
\usepackage{xspace}
\newcolumntype{C}[1]{>{\centering\let\newline\arraybackslash\hspace{0pt}}m{#1}}
\usepackage{graphicx} 
\usepackage{multirow}
\usepackage{booktabs}
\usepackage{float}
\usepackage{amsmath,amssymb,amsthm,amsfonts,xcolor, color,stmaryrd}
\usepackage{enumerate}
\usepackage[utf8]{inputenc}
\usepackage[english]{babel}
\usepackage{tikz}
\usetikzlibrary{matrix, shapes, arrows, positioning, chains}
\usepackage{ragged2e}
\usepackage{etoolbox}
\usepackage{lipsum}
\usepackage{cite}
\usepackage{hyperref}
\usepackage{siunitx}
\usepackage{authblk}
\usepackage[margin=0.9in]{geometry}
\usepackage[english]{babel}
\date{}
\usepackage{textgreek} 
\apptocmd{\frame}{}{\justifying}{} 

\title{Observational viability of Herglotz $f(R,T)$ gravity: A multi-probe Bayesian analysis}

\author[1]{Vishal M C \thanks{vishal.mc2023@vitstudent.ac.in}} 
\author[2]{Sankarsan Tarai \thanks{sankarsan.tarai@vit.ac.in}} 

\affil[1]{\it Department of Physics, School of Advanced Sciences, Vellore Institute of Technology, Chennai-600127, India} 

\affil[2]{\it Department of Mathematics, School of Advanced Sciences, Vellore Institute of Technology, Chennai-600127, India}

\begin{document}

\maketitle

\begin{abstract}
We investigate the observational viability of the linear Herglotz-type
$f(R,T)$ gravity model, $f(R,T)=R+\alpha T$, which incorporates both
geometry--matter coupling and non-conservative gravitational dynamics.
Unlike previous analyses\cite{Wazny:2025jth} based on illustrative parameter choices, we
perform a systematic Bayesian estimation of the four-dimensional
parameter space $\{H_0,A,w,\Phi_0\}$, where $A$ characterizes the
matter--geometry coupling, $w$ is the effective equation-of-state
parameter, and $\Phi_0$ denotes the present-day Herglotz field. The
background evolution is obtained by numerically integrating the coupled
Herglotz cosmological equations at every point in the parameter space.
We employ Cosmic Chronometer (CC), DESI DR2 baryon acoustic oscillation
(BAO), and Union3 Type-Ia supernova data, both independently and in
combination. The joint analysis yields
$H_0=66.67^{+1.32}_{-1.28}\,\mathrm{km\,s^{-1}\,Mpc^{-1}}$,
$A=1.57^{+0.66}_{-0.51}$, $w=-0.992^{+0.136}_{-0.119}$, and
$\Phi_0=-0.062^{+0.185}_{-0.157}$ at $68\%$ credibility. The
reconstructed Hubble expansion closely follows the flat $\Lambda$CDM
prediction over the redshift range probed by the CC data. However, the
deceleration parameter, effective equation of state, $Om(z)$ diagnostic,
and statefinder $\{r,s\}$ trajectories exhibit appreciable departures
from the concordance model, with the magnitude and redshift evolution
depending on the observational dataset. These results demonstrate that
Herglotz-type $f(R,T)$ gravity can provide an observationally viable
description of the late-time expansion while retaining distinguishable
cosmological signatures beyond the background Hubble history.
\end{abstract}

\section{Introduction}

The discovery of the accelerated expansion of the Universe has
provided one of the most important challenges to modern
cosmology. Within the framework of general relativity, the
observed late-time acceleration is usually attributed to a
cosmological constant, leading to the spatially flat
$\Lambda$CDM model. Although $\Lambda$CDM provides an excellent
description of a wide range of cosmological observations, the
nature of the cosmological constant and the origin of the observed
late-time acceleration remain open questions. In addition, the
cosmological constant is associated with well-known theoretical
issues, including the fine-tuning and coincidence problems\cite{Copeland:2006wr}.
These difficulties have motivated the investigation of modified
theories of gravity in which the accelerated expansion can emerge
from modifications of the gravitational sector rather than from a
fundamental cosmological constant.

Among the different extensions of general relativity, theories
with a nonminimal coupling between geometry and matter have
attracted considerable attention. In this context, $f(R,T)$
gravity, introduced by Harko et al.~\cite{Harko:2011kv}, generalizes
the Einstein--Hilbert action by allowing the gravitational
Lagrangian to depend explicitly on both the Ricci scalar $R$ and
the trace of the energy-momentum tensor,
\begin{equation}
    f=f(R,T).
\end{equation}
The explicit dependence on $T$ establishes a direct coupling
between the geometry and the matter sector and can consequently
modify both the background cosmological evolution and the
conservation properties of the energy-momentum tensor. A variety
of cosmological, astrophysical, and theoretical aspects of
$f(R,T)$ gravity have subsequently been investigated
\cite{Tretyakov:2018yph,Alvarenga:2013syu,Harko:2014gwa,Farias:2021jdz},
including its implications for cosmological expansion,
perturbations, compact objects, and gravitational phenomena.

A related but conceptually different direction arises when the
variational principle itself is generalized to describe
non-conservative systems. The standard Hamiltonian variational
principle is naturally formulated for conservative dynamics,
whereas dissipative and irreversible phenomena require a more
general framework. The Herglotz variational principle, originally
introduced to describe systems with friction, provides such a
generalization by allowing the Lagrangian to depend explicitly on
the action variable. Its covariant formulation makes it possible
to extend this idea to gravitational theories and thereby
incorporate non-conservative effects into the gravitational
variational structure. The resulting framework introduces an
additional Herglotz one-form, which acts as a background
contribution to the gravitational field equations. In a
homogeneous and isotropic cosmological setting, this one-form
reduces to a time-dependent function, thereby modifying the
background expansion dynamics. The Herglotz formalism has also
been applied to non-conservative gravity and related
cosmological scenarios, where connections with dissipative
cosmology and effective bulk-viscous behaviour have been
discussed
\cite{Lazo:2017udy,Paiva:2021kuk,Fabris:2017msx}.

The combination of the Herglotz variational principle with
$f(R,T)$ gravity provides a particularly interesting framework
because it simultaneously incorporates the geometry--matter
coupling characteristic of $f(R,T)$ theories and the
non-conservative structure associated with the Herglotz
formalism. Recently, Wazny et al.~\cite{Wazny:2025jth} developed a
Herglotz-type extension of $f(R,T)$ gravity and derived the
corresponding covariant gravitational field equations. Their
formulation contains additional contributions from the Herglotz
one-form and naturally recovers standard $f(R,T)$ gravity in the
appropriate limit. The authors also investigated the
non-conservation properties of the theory and showed that
energy-momentum conservation can be recovered under suitable
conditions on the Herglotz field. Beyond the cosmological
background, the Newtonian limit, motion of massive particles,
perihelion precession, and light deflection were also considered,
providing constraints on the Herglotz contribution from
gravitational phenomena.\\

The cosmological implications of Herglotz-type $f(R,T)$ gravity
were subsequently explored for a spatially flat FLRW universe
filled with a dust fluid. In particular, two functional forms,
$f(R,T)=R+\alpha T$ and $f(R,T)=R+\alpha T^{-1}$, were examined\cite{Wazny:2025jth}.
Because the Herglotz field entering the homogeneous cosmological
equations is non-dynamical, the resulting background system is
underdetermined and requires an additional closure relation. A
linear effective equation of state,
\begin{equation}
    p_{\rm eff}=w\rho_{\rm eff},
\end{equation}
was therefore introduced to close the system. For the linear
model $f(R,T)=R+\alpha T$, numerical solutions were shown to
produce an accelerating cosmological expansion and to provide a
reasonable description of a limited set of cosmic-chronometer
measurements for suitable choices of the model parameters.
Interestingly, the Herglotz contribution allows this linear
model to exhibit a richer cosmological evolution than the
corresponding standard $f(R,T)=R+\alpha T$ theory.

Despite these encouraging results, the cosmological analysis of
Herglotz-type $f(R,T)$ gravity remains relatively limited. In the
reference study, the cosmological parameters were selected to
illustrate the viability of the model rather than being obtained
from a comprehensive statistical parameter estimation. Moreover,
the cosmological comparison was primarily based on Hubble
expansion-rate measurements. The authors pointed out that a detailed comparison with a wider range of observational datasets is necessary before the cosmological viability of Herglotz-type $f(R,T)$ gravity can be assessed more completely. This leaves an important opportunity to investigate whether the apparent viability of the model persists when its free parameters are constrained simultaneously by independent cosmological probes.

Observational constraints are particularly important in this
context because the Herglotz-type model contains several
parameters whose effects on the background expansion can be
partially degenerate. In the linear model considered here, the
dimensionless coupling $A$, the effective equation-of-state
parameter $w$, and the present value of the Herglotz field
$\Phi_0$ determine the evolution of the dimensionless Hubble
parameter, while $H_0$ sets its present-day normalization.
Consequently, relying on a single observational probe may not be
sufficient to determine the allowed parameter space reliably.
Combining complementary probes that respond differently to the
expansion history can provide substantially stronger constraints
and can reveal whether the parameter preferences obtained from
individual datasets are mutually compatible.

In this work, we therefore perform a systematic observational
investigation of the linear Herglotz-type model
\begin{equation}
    f(R,T)=R+\alpha T.
\end{equation}
Rather than adopting fixed illustrative parameter values, we
treat the parameter set
\begin{equation}
    \boldsymbol{\theta}
    =
    \left\{
        H_0,A,w,\Phi_0
    \right\}
\end{equation}
as free parameters and determine their posterior distributions
using Bayesian parameter estimation. The background cosmological
equations are solved numerically for every point in the parameter
space, thereby allowing the Hubble function $H(z)$ to be obtained
directly from the underlying Herglotz cosmological equations
without introducing an independent phenomenological ansatz for
the expansion history.

To obtain complementary constraints, we consider three
independent observational probes: Cosmic Chronometer (CC)
measurements of the Hubble expansion rate, baryon acoustic
oscillation (BAO) measurements from the DESI DR2 release, and the
Union3 Type-Ia supernova compilation. Each dataset is first used
independently to determine how the inferred parameter space
depends on the observational probe, followed by a joint analysis
in which the three likelihoods are combined. This approach allows
us to investigate the extent to which the different probes
constrain or break the degeneracies among $H_0$, $A$, $w$, and
$\Phi_0$.

Beyond constraining the model parameters, we investigate the
resulting cosmological evolution through a set of complementary
diagnostics. In particular, we examine the the statefinder $\{r,s\}$ pair, the effective equation-of-state parameter $\omega_{\rm eff}(z)$, the deceleration parameter
$q(z)$ and the $Om(z)$ diagnostic. For comparison, we adopt a spatially flat $\Lambda$CDM model as the
reference cosmology, consistent with the standard cosmological framework supported by CMB observations \cite{Planck:2018vyg}. These diagnostics provide information beyond the
direct comparison of $H(z)$ and allow us to quantify departures
from the standard $\Lambda$CDM expansion history at different redshifts. The statefinder analysis is especially useful for identifying differences in higher-order derivatives of the expansion history that may remain hidden in a direct $H(z)$
comparison.

The main objective of this work is therefore twofold: first, we
provide a systematic multi-probe Bayesian constraints on the free parameters of the linear Herglotz-type $f(R,T)=R+\alpha T$ cosmological model considered here; and second, to determine whether the model remains phenomenologically
viable when its observationally constrained background evolution is examined through a broader set of cosmological diagnostics.
The results provide a quantitative assessment of the parameter space of the model and clarify the extent to which its apparent agreement with $\Lambda$CDM depends on the observational dataset
and on the particular cosmological quantity used for comparison.

The remainder of this paper is organized as follows. In Sec.~\ref{subsec:Cosmological Model}, we review the Herglotz-type $f(R,T)$
framework and derive the background equations for the linear model $f(R,T)=R+\alpha T$, including the effective equation-of-state closure and the dimensionless redshift formulation used in our numerical analysis. The observational constraints obtained from the CC, DESI DR2 BAO, Union3, and combined datasets are presented in Sec.~\ref{subsec:Observational Constraints}. Sec.~\ref{subsec:Cosmological Diagnostics}
investigates the reconstructed cosmological diagnostics, including $q(z)$, $\omega_{\rm eff}(z)$, $Om(z)$, and the statefinder $\{r,s\}$ plane. Finally, our main results and their
cosmological implications are summarized in
Sec.~\ref{subsec:Conclusion}.

\section{Cosmological Model}
\label{subsec:Cosmological Model}

In this section, we briefly review the Herglotz-type extension of
$f(R,T)$ gravity and derive the background cosmological equations
required for the observational analysis. The Herglotz variational
principle provides a generalized framework for describing
non-conservative systems by allowing the Lagrangian to depend
explicitly on an action variable. When this construction is
incorporated into modified gravity, additional contributions
associated with the Herglotz field appear in the gravitational
field equations. The resulting framework can therefore describe
cosmological scenarios in which the gravitational sector is not
strictly conservative. The formulation adopted here follows the
Herglotz-type $f(R,T)$ theory developed in Ref.~\cite{Wazny:2025jth}.

\subsection{Herglotz-type $f(R,T)$ gravity}

The Herglotz variational principle generalizes the conventional
Euler--Lagrange formulation by allowing the Lagrangian to depend
explicitly on the action variable. For a mechanical system, the
action variable $S$ satisfies
\begin{equation}
    \dot{S}=L(q,\dot q,S,t),
\end{equation}
where $q$ represents the generalized coordinate. Variation of this
action leads to the generalized Euler--Lagrange equation
\begin{equation}
    \frac{\partial L}{\partial q}
    -
    \frac{d}{dt}
    \left(
        \frac{\partial L}{\partial \dot q}
    \right)
    +
    \gamma
    \frac{\partial L}{\partial \dot q}
    =0,
    \qquad
    \gamma\equiv\frac{\partial L}{\partial S}.
\end{equation}
The additional term proportional to $\gamma$ is responsible for the
non-conservative contribution characteristic of the Herglotz
formalism. The field-theoretic generalization introduces an action
density $s^\mu$, satisfying
\begin{equation}
    \partial_\mu s^\mu
    =
    \mathcal{L}
    \left(
        \varphi,
        \partial_\mu\varphi,
        s^\mu,
        x^\mu
    \right),
\end{equation}
where $\varphi$ denotes the relevant field variables. The
corresponding generalized Euler--Lagrange equations acquire an
additional contribution involving the Herglotz vector
$\lambda^\mu$.

For the Herglotz-type extension of $f(R,T)$ gravity, the
gravitational field equations contain both the usual $f(R,T)$
terms and additional contributions associated with the Herglotz
field. In the notation of Ref.~\cite{Wazny:2025jth}, the generalized
field equations can be written schematically as
\begin{equation}
\begin{split}
f_R R_{\mu\nu}
-\frac{1}{2}f g_{\mu\nu}
+
\left(
g_{\mu\nu}\Box
-\nabla_\mu\nabla_\nu
\right)f_R
+
\mathcal{H}_{\mu\nu}
=
\frac{F}{2}T_{\mu\nu}
-
f_T
\left(
T_{\mu\nu}+\Theta_{\mu\nu}
\right),
\end{split}
\end{equation}
where
\begin{equation}
    f_R\equiv\frac{\partial f}{\partial R},
    \qquad
    f_T\equiv\frac{\partial f}{\partial T},
\end{equation}
and $\mathcal{H}_{\mu\nu}$ contains the additional Herglotz
contributions. The formalism reduces to the usual $f(R,T)$
framework when the Herglotz field vanishes, while the choice
$f(R,T)=R$ gives the corresponding non-conservative gravitational
theory. The Herglotz field is therefore an additional geometric
contribution to the gravitational dynamics.

\subsection{FLRW background}

We consider a spatially flat Friedmann--Lema\^{i}tre--Robertson--Walker
(FLRW) spacetime described by
\begin{equation}
    ds^2
    =
    -dt^2
    +
    a^2(t)
    \left(
        dx^2+dy^2+dz^2
    \right),
\end{equation}
where $a(t)$ denotes the scale factor and
\begin{equation}
    H\equiv\frac{\dot a}{a}
\end{equation}
is the Hubble parameter.

By spatial homogeneity and isotropy, the Herglotz vector is
restricted to the form
\begin{equation}
    \lambda_\mu
    =
    \left(
        \phi(t),0,0,0
    \right),
\end{equation}
where $\phi(t)$ represents the time-dependent Herglotz
contribution. The matter sector is described by a perfect fluid,
\begin{equation}
    T_{\mu\nu}
    =
    (\rho+p)u_\mu u_\nu
    +
    p g_{\mu\nu},
\end{equation}
where $\rho$ and $p$ are the energy density and pressure,
respectively, and
\begin{equation}
    u_\mu u^\mu=-1.
\end{equation}

Following Ref.~\cite{Wazny:2025jth}, we adopt a dust-dominated
background,
\begin{equation}
    p=0,
\end{equation}
so that the trace of the energy-momentum tensor becomes
\begin{equation}
    T=-\rho.
\end{equation}
For this homogeneous and isotropic background, the generalized
Friedmann equations of Herglotz-type $f(R,T)$ gravity reduce to
the modified cosmological equations given below. 

\subsection{Linear Model: $f(R,T)=R+\alpha T$}

We consider the simplest linear form of the Herglotz-type
$f(R,T)$ function,
\begin{equation}
    \boxed{
    f(R,T)=R+\alpha T,
    }
    \label{eq:fRT_model}
\end{equation}
where $\alpha$ is a constant coupling parameter describing the
strength of the $R$--$T$ interaction.

For this model,
\begin{equation}
    f_R=1,
    \qquad
    f_T=\alpha.
\end{equation}
Since $f_R$ is constant, all derivative terms involving $f_R$
vanish. The two independent FLRW equations then become
\begin{align}
    3H^2
    &=
    3\phi H
    +
    \frac{1}{2}
    (16\pi+3\alpha)\rho,
    \label{eq:Friedmann1}
    \\
    2\dot H+3H^2
    &=
    \dot\phi
    +
    2\phi H
    -
    \phi^2
    +
    \frac{\alpha}{2}\rho.
    \label{eq:Friedmann2}
\end{align}
These correspond to Eqs.~(119)--(120) of the reference model.

The first of these equations is an algebraic constraint on the
matter density. Solving Eq.~\eqref{eq:Friedmann1} for $\rho$ gives
\begin{equation}
    \rho
    =
    \frac{2}{16\pi+3\alpha}
    \left(
        3H^2-3\phi H
    \right).
    \label{eq:rho_model}
\end{equation}
Substituting this relation into the second Friedmann equation
yields
\begin{equation}
    2\dot H-\dot\phi
    =
    2\phi H
    -
    3H^2
    -
    \phi^2
    +
    \frac{3\alpha}{16\pi+3\alpha}
    \left(
        H^2-\phi H
    \right).
    \label{eq:dynamical_Hphi}
\end{equation}
Thus, the cosmological dynamics are governed by the two unknown
functions $H(t)$ and $\phi(t)$. As emphasized in the reference
model, an additional relation is therefore required to close the
system.

\subsection{Effective Equation of State}

To close the cosmological system, we adopt the effective
equation-of-state prescription
\begin{equation}
    p_{\rm eff}
    =
    w\rho_{\rm eff},
    \label{eq:effective_eos}
\end{equation}
where $w$ is a constant effective equation-of-state parameter.
This prescription is introduced at the level of the effective
geometrical sector and should therefore not be confused with the
pressure of the physical dust component, for which $p=0$.

The effective density and pressure associated with the modified
gravitational sector can be introduced through
\begin{align}
    3H^2
    &=
    8\pi
    \left(
        \rho+\rho_{\rm eff}
    \right),
    \\
    2\dot H+3H^2
    &=
    -8\pi
    \left(
        p+p_{\rm eff}
    \right).
\end{align}
For the linear model considered here, the effective equation of
state provides the additional relation required to determine the
evolution of both $H$ and $\phi$. The reference paper explicitly
introduces this closure because the cosmological system contains
two unknown functions but only one independent dynamical equation.

Using the effective equation-of-state relation together with the
linear $f(R,T)$ field equations gives the evolution equation for
the Herglotz field,
\begin{equation}
\begin{split}
    \dot\phi
    &=
    \phi^2
    -
    2\phi H
    -
    \frac{\alpha}{2}\rho
    -
    3w
    \left(
        \phi H+\frac{\alpha}{2}\rho
    \right).
\end{split}
\label{eq:phi_dot}
\end{equation}
After eliminating the matter density using the algebraic
constraint, this equation becomes
\begin{equation}
\begin{split}
    \dot\phi
    &=
    \phi^2
    -
    \frac{
        (32\pi+3\alpha)\phi
        +3\alpha H
    }{
        16\pi+3\alpha
    }H
    -
    3w
    \frac{
        16\pi\phi+3\alpha H
    }{
        16\pi+3\alpha
    }H.
\end{split}
\label{eq:phi_dot_reduced}
\end{equation}
This is the closure relation used to obtain the numerical
background evolution of the model.

\subsection{Dimensionless Formulation}

For numerical calculations, it is convenient to introduce the
dimensionless variables
\begin{equation}
    \tau=H_0t,
    \qquad
    H=H_0h,
    \qquad
    \rho=\frac{3H_0^2}{8\pi}r,
    \qquad
    \phi=H_0\Phi,
    \qquad
    \alpha=\frac{16\pi}{3}A,
    \label{eq:dimensionless_variables}
\end{equation}
where $H_0$ denotes the present-day Hubble constant. The
dimensionless coupling $A$ is therefore related to the original
coupling constant $\alpha$ through
\begin{equation}
    A=\frac{3\alpha}{16\pi}.
\end{equation}
These definitions are those adopted in the reference model.

In terms of these variables, Eq.~\eqref{eq:dynamical_Hphi}
takes the form
\begin{equation}
    2\frac{dh}{d\tau}
    -
    \frac{d\Phi}{d\tau}
    =
    \frac{2+A}{1+A}\Phi h
    -
    \Phi^2
    -
    \frac{3+2A}{1+A}h^2.
    \label{eq:dimensionless_dynamical}
\end{equation}
The corresponding algebraic constraint becomes
\begin{equation}
    h^2
    =
    \Phi h
    +
    (1+A)r,
    \label{eq:dimensionless_constraint}
\end{equation}
and consequently the dimensionless matter density can be
reconstructed as
\begin{equation}
    r(z)
    =
    \frac{
        h^2(z)-\Phi(z)h(z)
    }{1+A}.
    \label{eq:dimensionless_density}
\end{equation}
Equation~\eqref{eq:dimensionless_density} is the same density
reconstruction subsequently used in the numerical analysis of the
reference model. 

\subsection{Background Evolution in Redshift}

The observational analysis is performed in terms of the
cosmological redshift rather than cosmic time. Using
\begin{equation}
    1+z=\frac{1}{a},
\end{equation}
the time derivative can be expressed as
\begin{equation}
    \frac{d}{d\tau}
    =
    -(1+z)h(z)\frac{d}{dz}.
    \label{eq:time_redshift}
\end{equation}

Combining the dimensionless equations with the effective
equation-of-state closure gives the coupled first-order system
\begin{equation}
\boxed{
\frac{d\Phi}{dz}
=
\frac{1}{1+z}
\left[
-\frac{\Phi^2}{h}
+
3w
\frac{\Phi+Ah}{1+A}
+
\frac{2+A}{1+A}\Phi
+
\frac{A}{1+A}h
\right],
}
\label{eq:Phi_z}
\end{equation}
and
\begin{equation}
\boxed{
\frac{dh}{dz}
=
\frac{3}{2(1+z)}
\left[
h+
w\frac{\Phi+Ah}{1+A}
\right].
}
\label{eq:h_z}
\end{equation}
These equations correspond to Eqs.~(138)--(139) of the reference
model and constitute the background system that we numerically
integrate in the present analysis.

The present-day normalization of the dimensionless Hubble
parameter is imposed as
\begin{equation}
    h(0)=1,
    \label{eq:h_initial}
\end{equation}
while the present-day value of the Herglotz field is specified by
\begin{equation}
    \Phi(0)=\Phi_0.
    \label{eq:Phi_initial}
\end{equation}
The physical Hubble parameter is then reconstructed from
\begin{equation}
    \boxed{
    H(z)=H_0h(z).
    }
    \label{eq:H_reconstruction}
\end{equation}

The numerical solution of Eqs.~\eqref{eq:Phi_z} and
\eqref{eq:h_z}, together with the initial conditions
\eqref{eq:h_initial} and \eqref{eq:Phi_initial}, therefore provides
the complete background expansion history required for the
observational analysis. The original study solved these equations
numerically using selected values of the model parameters and
demonstrated agreement with the available cosmic-chronometer data
over approximately $z\lesssim2$. 

\section{Observational Constraints}
\label{subsec:Observational Constraints}

\subsection{Free Parameters and Observational Analysis}

The reference study considered illustrative parameter choices for
$A$, $w$, and $H_0$, while exploring the effect of different values
of the present-day Herglotz field $\Phi_0$. The authors also noted that
these parameter choices were not intended to represent an optimal
observational determination of the model parameters\cite{Wazny:2025jth}. In the present work, we instead perform a systematic observational
constraint of the model by allowing all relevant cosmological
parameters to vary simultaneously.

Accordingly, the parameters
\begin{equation}
    H_0,\qquad A,\qquad w,\qquad \Phi_0
\end{equation}
are treated as free parameters, and the cosmological parameter vector
is defined as
\begin{equation}
    \boxed{
    \boldsymbol{\theta}
    =
    \left\{
        H_0,\,
        A,\,
        w,\,
        \Phi_0
    \right\}.
    }
    \label{eq:parameter_vector}
\end{equation}

For every point in this four-dimensional parameter space, the coupled
background equations, Eqs.~\eqref{eq:Phi_z} and \eqref{eq:h_z}, are
numerically integrated to obtain the dimensionless Hubble parameter
$h(z)$ and the Herglotz field $\Phi(z)$. The physical Hubble parameter
is subsequently reconstructed through Eq.~\eqref{eq:H_reconstruction}.
This numerical expansion history is then used consistently in the
construction of the observational likelihoods.

To constrain the model parameters, we consider three complementary
observational probes: Cosmic Chronometer (CC) measurements, DESI DR2
baryon acoustic oscillation (BAO) measurements, and the Union3
Type-Ia supernova compilation. Each dataset is first analyzed
independently in order to investigate its individual constraining
power, followed by a joint analysis combining all three probes.
The posterior distributions of the free parameters are obtained
using Bayesian parameter estimation through Markov Chain Monte Carlo\cite{Foreman-Mackey:2012any,Lewis:2019xzd}
(MCMC).

The corresponding observational datasets, likelihood functions and adopted priors are described in the following sections, followed by the resulting parameter constraints and
cosmological diagnostics.

\subsection{Cosmic Chronometer Data}
\label{subsec:Cosmic Chronometer}

The Cosmic Chronometer (CC) dataset provides a direct and
model-independent method for constraining the Hubble expansion
rate $H(z)$ at different redshifts. In this work, we use a compilation of 31 measurements of the
Hubble expansion rate covering the redshift range $0.07\lesssim z\lesssim2.34$\cite{Guo:2015gpa}. The cosmic chronometer
method is based on estimating the differential ages of passively
evolving, massive galaxies at different redshifts. Since the
redshift evolution of the cosmic scale factor is related to cosmic
time through
\begin{equation}
    H(z)=-\frac{1}{1+z}\frac{dz}{dt},
\end{equation}
the differential age evolution of these galaxies provides a direct
estimate of the Hubble parameter \cite{Jimenez:2001gg}. In
particular, the method relies on the determination of the age
difference between two nearby galaxy populations, allowing
$dt/dz$ and hence $H(z)$ to be inferred observationally. An
important advantage of the CC approach is that the measurements
directly probe the expansion rate without requiring a specific
cosmological model or a prior parametrization of $H(z)$. Therefore,
the CC dataset provides a useful and relatively direct probe of the
background expansion history and is particularly suitable for
constraining the numerically reconstructed $H(z)$ of the Herglotz
Model.

To constrain the model parameters using the CC measurements, we
construct the corresponding chi-square statistic as
\begin{equation}
    \chi^2_{H(z)}
    =
    \sum_{i=1}^{31}
    \frac{\left[H(\boldsymbol{\theta},z_i)
    -H_{\rm obs}(z_i)\right]^2}
    {\sigma^2(z_i)},
\end{equation}
where $H(\boldsymbol{\theta},z_i)$ denotes the theoretical Hubble
parameter evaluated at the redshift $z_i$ for a given set of model
parameters
\begin{equation}
    \boldsymbol{\theta}
    =
    \left\{H_0,A,w,\Phi_0\right\},
\end{equation}
while $H_{\rm obs}(z_i)$ and $\sigma(z_i)$ represent the observed
Hubble parameter and its corresponding $1\sigma$ uncertainty,
respectively. Since the CC measurements are treated as
statistically independent, the total chi-square is obtained as the
sum of the individual contributions from the 31 measurements. The
corresponding likelihood is then given by
\begin{equation}
    \ln\mathcal{L}_{\rm CC}
    =
    -\frac{1}{2}\chi^2_{H(z)}.
\end{equation}
This likelihood is evaluated for every point in the MCMC parameter
space using the numerically reconstructed Hubble expansion history
of the Herglotz Model.

\begin{figure}[htbp]
\centering
\includegraphics[width=0.55\linewidth]{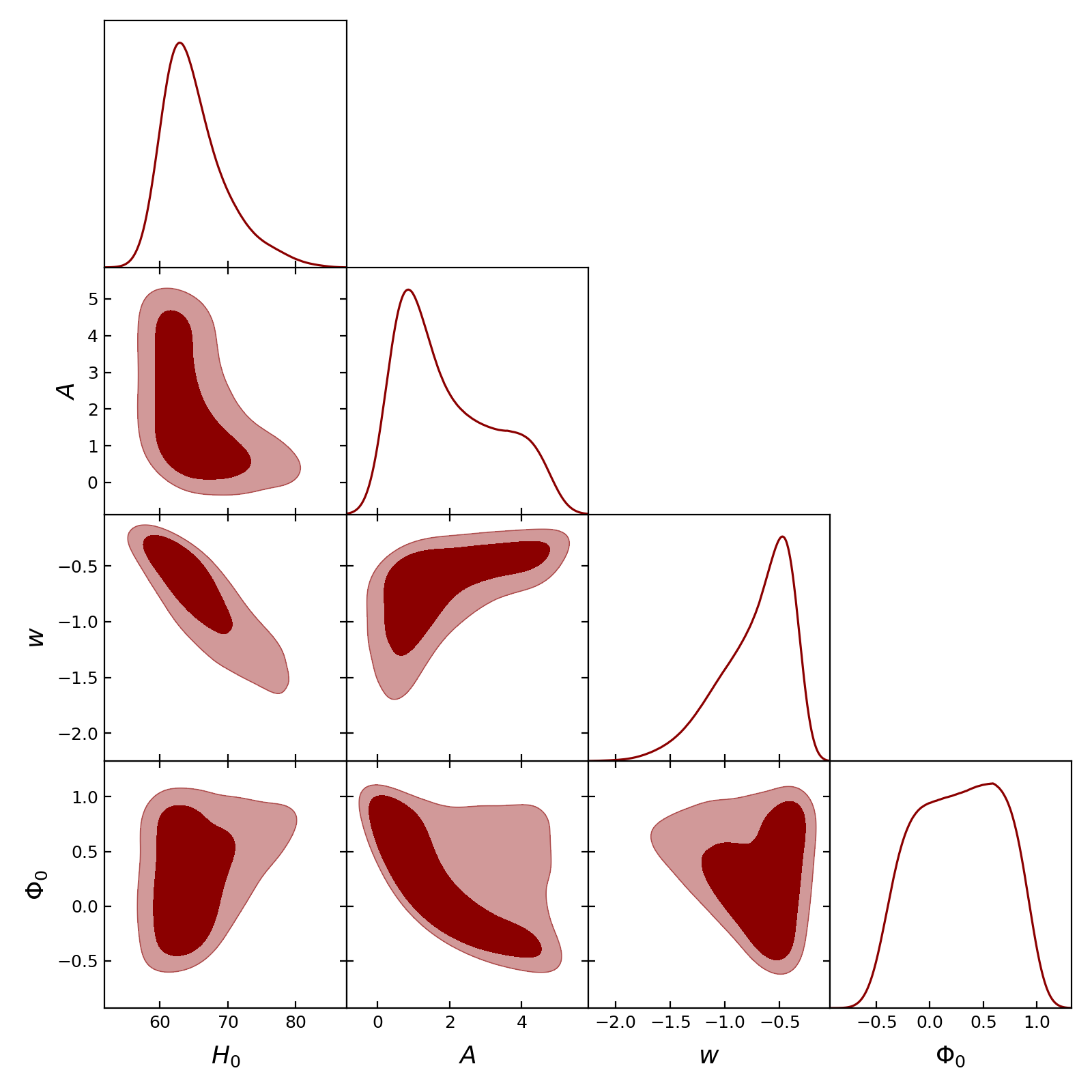}
\caption{Two-dimensional and marginalised one-dimensional posterior
distributions for the Herglotz Model obtained from the CC dataset.
The dark and light shaded regions denote the $68\%$ and $95\%$
credible regions, respectively, for the parameters
$(H_0,A,w,\Phi_0)$.}
    \label{CC_corner.png}
\end{figure}

The two-dimensional and marginalised one-dimensional posterior
distributions obtained from the MCMC analysis of the Herglotz
Model, $f(R,T)=R+\alpha T$, using the Cosmic Chronometer (CC)
dataset are presented in the triangle plot of
Fig.~\ref{CC_corner.png}. The dark- and light-shaded regions
correspond to the $68\%$ and $95\%$ credible regions, respectively,
for the four sampled parameters $(H_0,A,w,\Phi_0)$. The marginalised
posterior distribution of the Hubble constant is reasonably
concentrated, with a peak around
$H_0\simeq64$~km~s$^{-1}$~Mpc$^{-1}$ and an extended tail toward
higher values. This indicates that the CC dataset alone favours a
comparatively low value of $H_0$, although a broader range of values
remains statistically allowed. This is consistent with early-Universe/CMB-inferred values \cite{Planck:2018vyg} and in tension with local distance-ladder measurements \cite{Riess:2021jrx}. A noticeable negative correlation
between $H_0$ and $w$ is also visible in the corresponding
two-dimensional posterior, indicating that variations in these
parameters can be partially compensated while maintaining a
comparable description of the observed expansion history.

\subsection{Baryon Acoustic Oscillation Data}
\label{subsec:BAO}

Baryon Acoustic Oscillations (BAO) provide a powerful geometric
probe of the expansion history of the Universe by measuring the
characteristic scale imprinted in the spatial distribution of
matter. The BAO scale acts as a standard ruler and allows
cosmological distances to be constrained at different effective
redshifts. In this work, we use the BAO measurements from the second
data release (DR2) of the Dark Energy Spectroscopic Instrument
(DESI) \cite{DESI:2025zgx,DESI:2025zpo}. The DESI DR2 BAO compilation combines
measurements from different galaxy and quasar tracers together with
the high-redshift Ly$\alpha$ forest measurements, providing
constraints on the expansion history over a broad redshift range.\\

For the present analysis, we use the combined ``ALL\_GCcomb''
Gaussian BAO data vector, which contains 13 measurements of the
transverse, radial, and isotropic BAO distance indicators. The
observables are given in terms of
$D_M(z)/r_d$, $D_H(z)/r_d$, and $D_V(z)/r_d$, where $r_d$ denotes
the sound horizon at the drag epoch. The data vector covers the
effective redshift range $0.295\leq z_{\rm eff}\leq2.330$ and
contains measurements from the BGS, LRG, ELG, QSO, and Ly$\alpha$
tracers. The complete $13\times13$ covariance matrix, including the
correlations among the individual measurements, is used in the
likelihood analysis. The data vector and covariance matrix employed
in our analysis are taken from the DESI DR2 BAO data products used
in the Cobaya analysis framework.\\

For a spatially flat background, the transverse comoving distance
is calculated from the reconstructed Hubble expansion rate as
\begin{equation}
    D_M(z)
    =
    c\int_0^z\frac{dz'}{H(z')},
\end{equation}
while the radial Hubble distance is given by
\begin{equation}
    D_H(z)
    =
    \frac{c}{H(z)}.
\end{equation}
The angle-averaged distance measure entering the isotropic BAO
observable is defined as
\begin{equation}
    D_V(z)
    =
    \left[zD_M^2(z)D_H(z)\right]^{1/3}.
\end{equation}
The theoretical quantities entering the likelihood are therefore
constructed as
\begin{equation}
    \frac{D_M(z)}{r_d},
    \qquad
    \frac{D_H(z)}{r_d},
    \qquad
    \frac{D_V(z)}{r_d}.
\end{equation}

Since the Herglotz Model does not possess a closed-form expression
for $H(z)$ under the effective equation-of-state closure adopted
here, the coupled background equations for $h(z)$ and $\Phi(z)$ are
numerically integrated for every point in the MCMC parameter space.
The resulting dimensionless Hubble parameter is used to reconstruct
the physical expansion rate according to
\begin{equation}
    H(z)=H_0h(z),
\end{equation}
which is subsequently employed in the calculation of both
$D_M(z)$ and $D_H(z)$. In particular, $D_M(z)$ is obtained through
the numerical integration of $1/H(z)$, whereas $D_H(z)$ is evaluated
directly from the reconstructed Hubble function. The corresponding
$D_V(z)$ is then obtained from the combination of these two distance
measures. This procedure ensures that the same numerically
reconstructed background expansion history is used consistently
throughout the BAO likelihood analysis. 

In the present analysis, the sound horizon at the drag epoch is
held fixed at\cite{Planck:2018vyg}
\begin{equation}
    r_d=147.05~{\rm Mpc},
\end{equation}
following the fiducial value adopted in the BAO implementation used
for the analysis. The four cosmological parameters
\begin{equation}
    \boldsymbol{\theta}
    =
    \left\{
    H_0,A,w,\Phi_0
    \right\}
\end{equation}
are varied simultaneously in the MCMC analysis, while $r_d$ is not
treated as an additional free parameter. 

The BAO likelihood is constructed using the full covariance matrix
rather than treating the 13 measurements as statistically
independent. Defining the residual vector as
\begin{equation}
    \Delta\mathbf{D}
    =
    \mathbf{D}_{\rm th}
    -
    \mathbf{D}_{\rm obs},
\end{equation}
the corresponding chi-square is
\begin{equation}
    \chi^2_{\rm BAO}
    =
    \Delta\mathbf{D}^{\,T}
    C_{\rm BAO}^{-1}
    \Delta\mathbf{D},
\end{equation}
where $C_{\rm BAO}$ denotes the full $13\times13$ covariance matrix.
The BAO contribution to the posterior is consequently obtained
from
\begin{equation}
    \ln\mathcal{L}_{\rm BAO}
    =
    -\frac{1}{2}\chi^2_{\rm BAO}.
\end{equation}
This treatment accounts for the correlations among the DESI DR2
BAO observables and provides a consistent constraint on the
four-dimensional Herglotz Model parameter space.

\begin{figure}[htbp]
\centering
\includegraphics[width=0.55\linewidth]{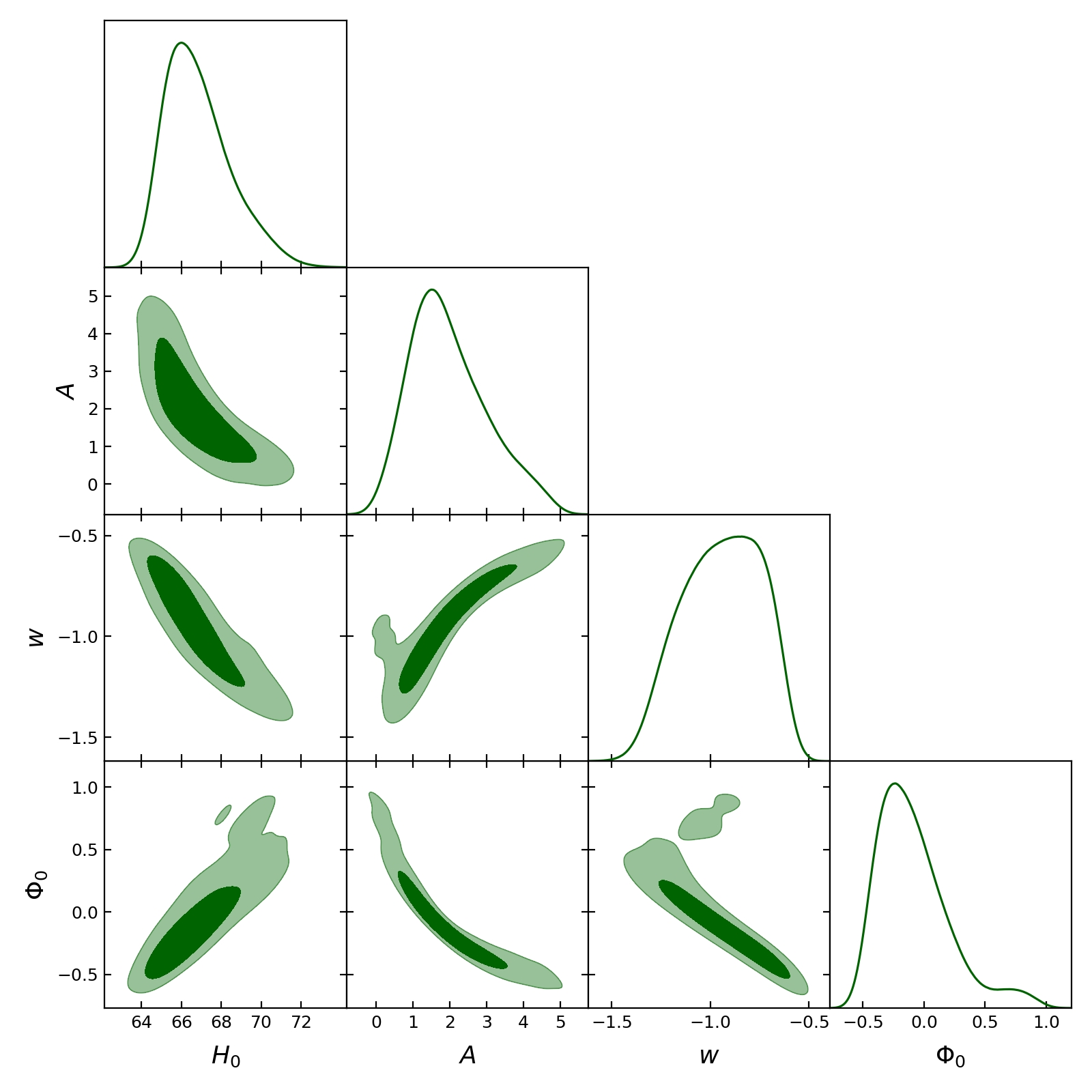}
\caption{Two-dimensional and marginalised one-dimensional posterior
distributions for the Herglotz Model obtained from the DESI DR2 BAO
dataset. The dark and light shaded regions denote the $68\%$ and
$95\%$ credible regions, respectively, for the parameters
$(H_0,A,w,\Phi_0)$.}
    \label{BAO_corner.png}
\end{figure}

\subsection{Union3 Supernova Data}
\label{subsec:Union3}

Type-Ia supernovae (SNe Ia) provide an important probe of the
late-time expansion history through their observed luminosity
distances. In this work, we use the Union3 supernova compilation
from the Supernova Cosmology Project, which provides a compressed
set of binned distance-modulus measurements together with their
statistical and systematic covariance information
\cite{Rubin:2023jdq}. The compressed Union3 data product contains the
redshift, distance modulus, and corresponding inverse covariance
matrix, allowing the correlations among the binned measurements to
be incorporated directly into the likelihood analysis. The full
inverse covariance matrix supplied with the dataset is used in our
analysis rather than treating the individual supernova measurements
as statistically independent. 

For a spatially flat background, the theoretical luminosity
distance is related to the Hubble expansion rate through the
comoving distance,
\begin{equation}
    D_C(z)
    =
    c\int_0^z\frac{dz'}{H(z')},
\end{equation}
where $c$ denotes the speed of light. The luminosity distance is
then given by
\begin{equation}
    D_L(z)
    =
    (1+z)D_C(z).
\end{equation}
The corresponding theoretical distance modulus is
\begin{equation}
    \mu_{\rm th}(z)
    =
    5\log_{10}
    \left[
    \frac{D_L(z)}{\mathrm{Mpc}}
    \right]
    +25.
\end{equation}

For the Herglotz Model considered here, the Hubble expansion history
is obtained by numerically integrating the coupled background
equations for $h(z)$ and $\Phi(z)$. For every point in the MCMC
parameter space, the resulting dimensionless Hubble parameter is
converted into the physical Hubble rate according to
\begin{equation}
    H(z)=H_0h(z).
\end{equation}
The reconstructed $H(z)$ is subsequently used to calculate the
comoving distance through the numerical integration of $1/H(z)$,
from which $D_L(z)$ and $\mu_{\rm th}(z)$ are obtained. Thus, the
same numerical background solution used in the CC and BAO analyses
is consistently propagated into the supernova likelihood. The
numerical implementation evaluates the luminosity distance using a
cumulative integral of $1/H(z)$ over a dense redshift grid.

To constrain the model parameters with the Union3 data, we define
the residual vector
\begin{equation}
    \Delta\boldsymbol{\mu}
    =
    \boldsymbol{\mu}_{\rm th}
    -
    \boldsymbol{\mu}_{\rm obs},
\end{equation}
where $\boldsymbol{\mu}_{\rm th}$ and
$\boldsymbol{\mu}_{\rm obs}$ denote the theoretical and observed
distance-modulus vectors, respectively. The corresponding
chi-square is constructed using the full inverse covariance matrix,
\begin{equation}
    \chi^2_{\rm Union3}
    =
    \Delta\boldsymbol{\mu}^{\,T}
    C_{\rm Union3}^{-1}
    \Delta\boldsymbol{\mu},
\end{equation}
where $C_{\rm Union3}$ denotes the covariance matrix containing both
statistical and systematic contributions. 
The corresponding logarithmic likelihood is therefore
\begin{equation}
    \ln\mathcal{L}_{\rm Union3}
    =
    -\frac{1}{2}\chi^2_{\rm Union3}.
\end{equation}
The four parameters
\begin{equation}
    \boldsymbol{\theta}
    =
    \left\{
    H_0,A,w,\Phi_0
    \right\}
\end{equation}
are varied simultaneously in the MCMC analysis using the same
parameter priors adopted for the CC and DESI DR2 BAO constraints.\\

\begin{figure}[htbp]
\centering
\includegraphics[width=0.55\linewidth]{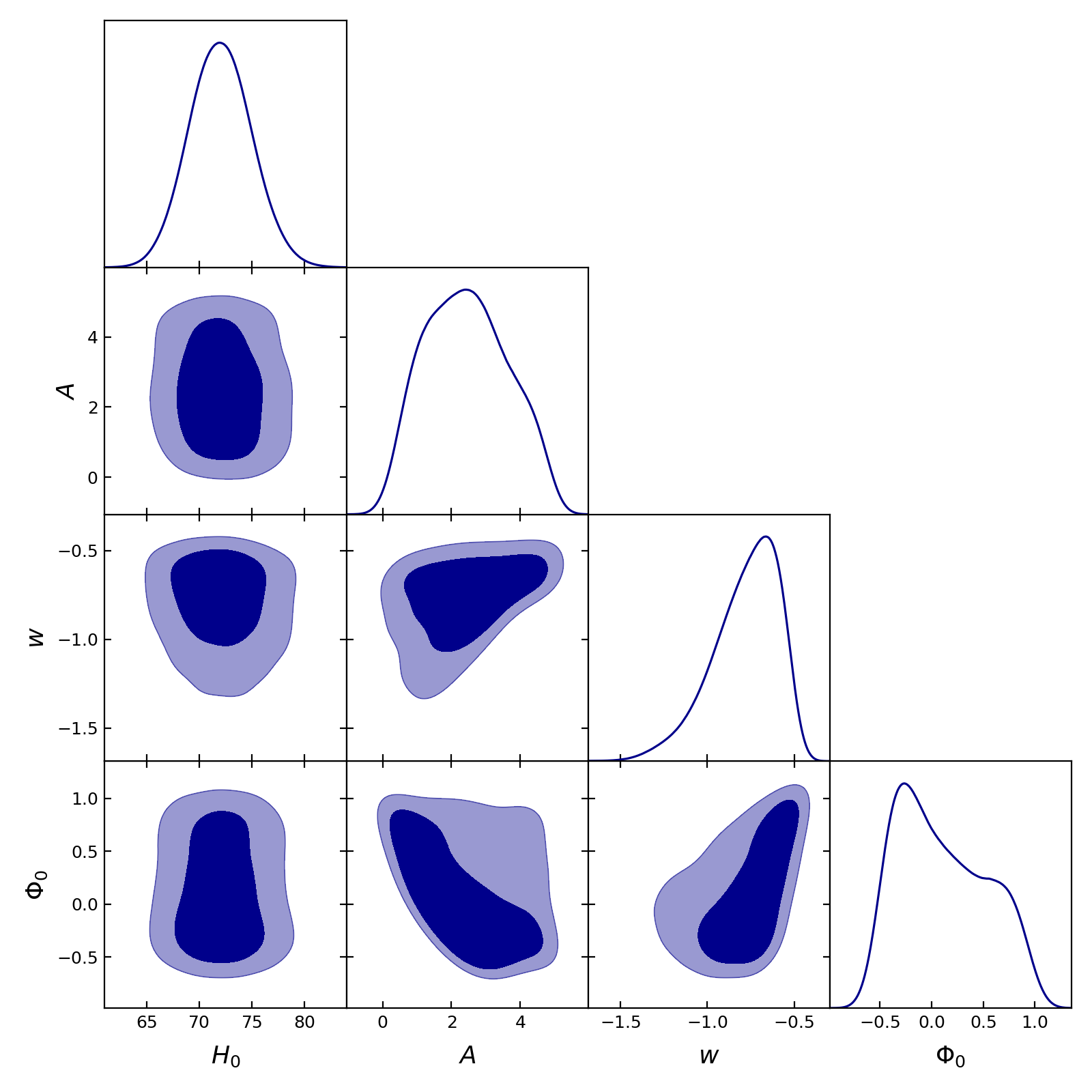}
\caption{Two-dimensional and marginalised one-dimensional posterior distributions for the parameters $(H_0,A,w,\Phi_0)$ of the Herglotz Model obtained from the Union3 supernova dataset. The dark and light shaded regions denote the $68\%$ and $95\%$ credible regions, respectively.}
    \label{Union3_corner.png}
\end{figure}

The resulting two-dimensional and marginalised one-dimensional
posterior distributions for the Herglotz Model obtained from the
Union3 dataset are shown in Fig.~\ref{Union3_corner.png}. The dark
and light shaded regions represent the $68\%$ and $95\%$ credible
regions, respectively. The marginalised posterior of $H_0$ is
shifted toward higher values compared with the CC and BAO analyses,
with a peak around $H_0\simeq72$~km~s$^{-1}$~Mpc$^{-1}$ and most of
the posterior probability concentrated approximately within
$68\lesssim H_0\lesssim76$~km~s$^{-1}$~Mpc$^{-1}$. The corresponding
two-dimensional contours show comparatively weak correlations
between $H_0$ and the remaining parameters, indicating that the
Union3 distance-modulus data constrain the overall expansion scale
with relatively limited degeneracy with the other model parameters.

The coupling parameter $A$ remains comparatively broad, with its
marginalised posterior extending toward higher values and exhibiting
a peak around $A\simeq2$--$3$. Its two-dimensional posterior with
$w$ displays an elongated and curved structure, indicating a
persistent degeneracy between the coupling strength and the
effective equation-of-state parameter. The posterior distribution
of $w$ is concentrated predominantly in the interval
$-1.3\lesssim w\lesssim-0.5$, with a peak around $w\simeq-0.7$.
Finally, the Herglotz-field parameter $\Phi_0$ remains relatively
weakly constrained, with a broad posterior extending over a
substantial fraction of the allowed prior range and a tail toward
positive values. The corresponding two-dimensional contours
indicate that $\Phi_0$ participates in parameter degeneracies with
$A$ and $w$. Overall, the Union3 analysis favours a comparatively
higher $H_0$ and leaves the coupling and Herglotz-field parameters
less tightly constrained, highlighting the complementary
constraining behaviour of supernova distance measurements relative
to the CC and BAO datasets.

\subsection{Combined Analysis}
\label{subsec:combined}

\begin{figure}[htbp]
\centering
\includegraphics[width=0.55\linewidth]{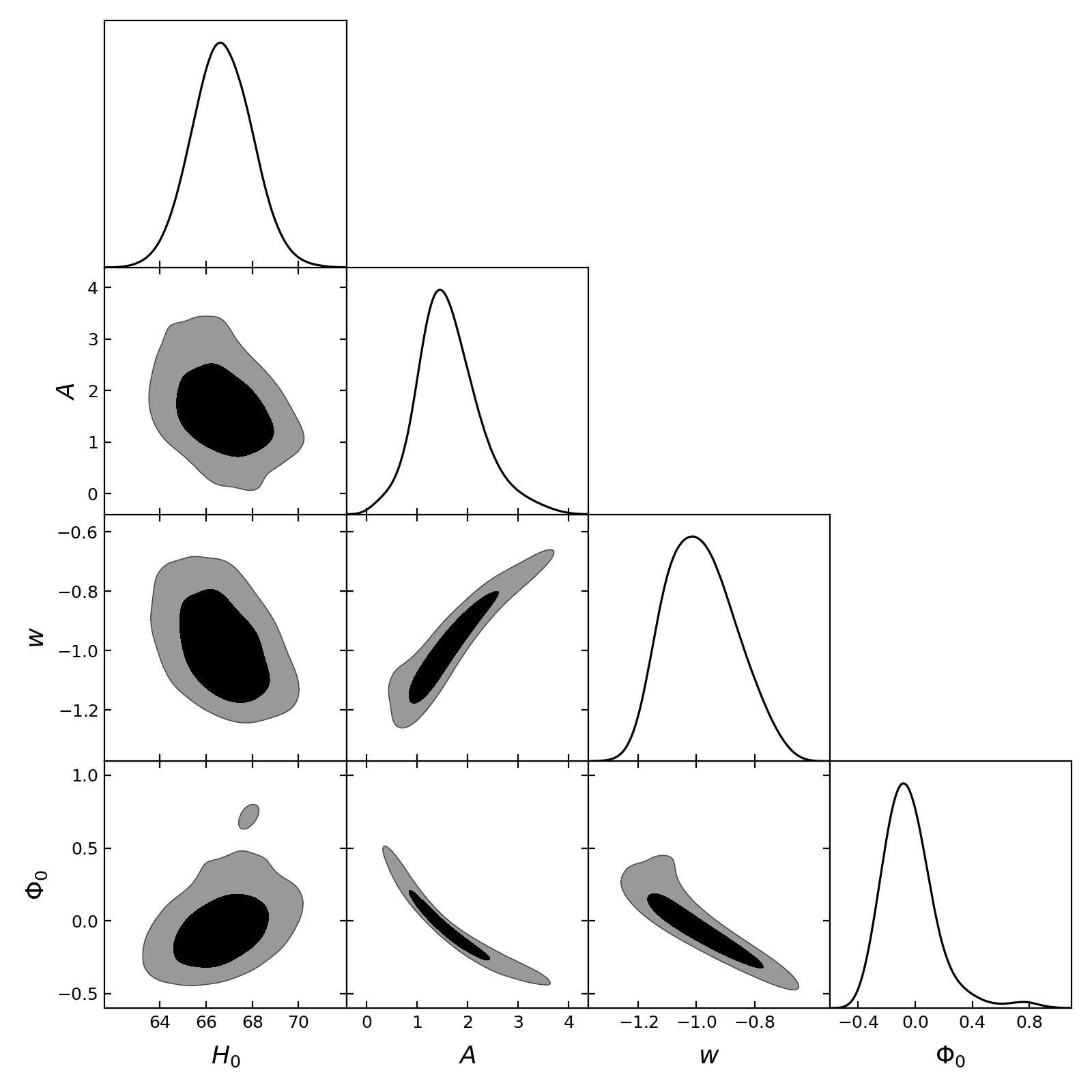}
\caption{Two-dimensional and marginalised one-dimensional posterior
distributions for the parameters $(H_0,A,w,\Phi_0)$ of the Herglotz
Model obtained from the combined CC+Union3+DESI DR2 BAO analysis.
The inner and outer contours correspond to the $68\%$ and $95\%$
credible regions, respectively.}
    \label{Combined_corner.png}
\end{figure}

The joint observational constraints obtained by combining the Cosmic
Chronometer (CC), Union3 supernova, and DESI DR2 BAO datasets are
presented in Fig.~\ref{Combined_corner.png}. In the combined analysis,
the four physical parameters of the Herglotz Model,
$(H_0,A,w,\Phi_0)$, are constrained simultaneously. The sound horizon
at the drag epoch, $r_d$, is fixed to the fiducial value adopted in
the BAO analysis and is therefore not treated as an additional free
parameter. For every MCMC sample, the coupled background equations
are numerically integrated to reconstruct the dimensionless Hubble
parameter $h(z)$, and the resulting expansion history is used
consistently in the CC, Union3, and BAO likelihoods.

Assuming statistical independence between the three observational
datasets, the total chi-square is given by
\begin{equation}
    \chi^2_{\rm tot}
    =
    \chi^2_{\rm CC}
    +
    \chi^2_{\rm Union3}
    +
    \chi^2_{\rm BAO},
\end{equation}
where $\chi^2_{\rm CC}$, $\chi^2_{\rm Union3}$, and
$\chi^2_{\rm BAO}$ denote the contributions from the cosmic
chronometer, supernova, and BAO datasets, respectively. Equivalently,
the total likelihood can be written as
\begin{equation}
    \ln\mathcal{L}_{\rm tot}
    =
    \ln\mathcal{L}_{\rm CC}
    +
    \ln\mathcal{L}_{\rm Union3}
    +
    \ln\mathcal{L}_{\rm BAO}.
\end{equation}
The posterior distribution is then sampled according to
\begin{equation}
    \ln\mathcal{P}(\boldsymbol{\theta}\mid\mathcal{D})
    =
    \ln\mathcal{L}_{\rm tot}
    +
    \ln\Pi(\boldsymbol{\theta}),
\end{equation}
where
\begin{equation}
    \boldsymbol{\theta}
    =
    (H_0,A,w,\Phi_0)
\end{equation}
denotes the four-dimensional parameter vector, $\mathcal{D}$ denotes
the combined observational dataset, and
$\Pi(\boldsymbol{\theta})$ represents the adopted prior distribution.
Thus, the joint analysis combines the complementary information from
the expansion-rate, supernova distance, and BAO measurements while
constraining the same four physical parameters of the Herglotz Model.

The marginalised posterior distributions and the corresponding
two-dimensional confidence contours are shown in
Fig.~\ref{Combined_corner.png}. The marginalised posterior of the
Hubble constant is relatively well localized, with a maximum
probability around
$H_0\simeq67~{\rm km\,s^{-1}\,Mpc^{-1}}$. Compared with the individual
datasets, the joint constraint substantially reduces the allowed
range of $H_0$, indicating that the combination of the three
observational probes provides a more precise determination of the
present-day expansion rate. The coupling parameter $A$ is also more
tightly constrained, with the posterior favouring values around
$A\simeq1$--$2$. A clear correlation between $A$ and $w$ remains
visible in the corresponding two-dimensional posterior, indicating
that these parameters can partially compensate each other in
determining the background expansion history. Nevertheless, the
joint analysis significantly reduces the extent of this degeneracy
relative to the individual-probe constraints.

The posterior distribution of the effective equation-of-state
parameter $w$ is concentrated around values close to $w\simeq-1$,
although a finite range of values around this region remains allowed.
The combined constraint therefore favours an effective equation of
state close to the $\Lambda$CDM value while still allowing deviations
associated with the Herglotz modification. The present-day Herglotz
field $\Phi_0$ is also more localized than in several of the
individual analyses, with its posterior concentrated around values
close to zero. Noticeable correlations between $\Phi_0$ and both
$A$ and $w$ remain evident in the two-dimensional contours. These
correlations reflect the compensating role of the Herglotz field and
the coupling parameter in reconstructing a similar background
expansion history. Overall, the combined posterior demonstrates that
the joint use of CC, Union3, and BAO data substantially reduces the
parameter degeneracies present in the individual analyses and yields
a comparatively well-constrained region in the
$(H_0,A,w,\Phi_0)$ parameter space.

\begin{figure}[htbp]
\centering
\includegraphics[width=0.55\linewidth]{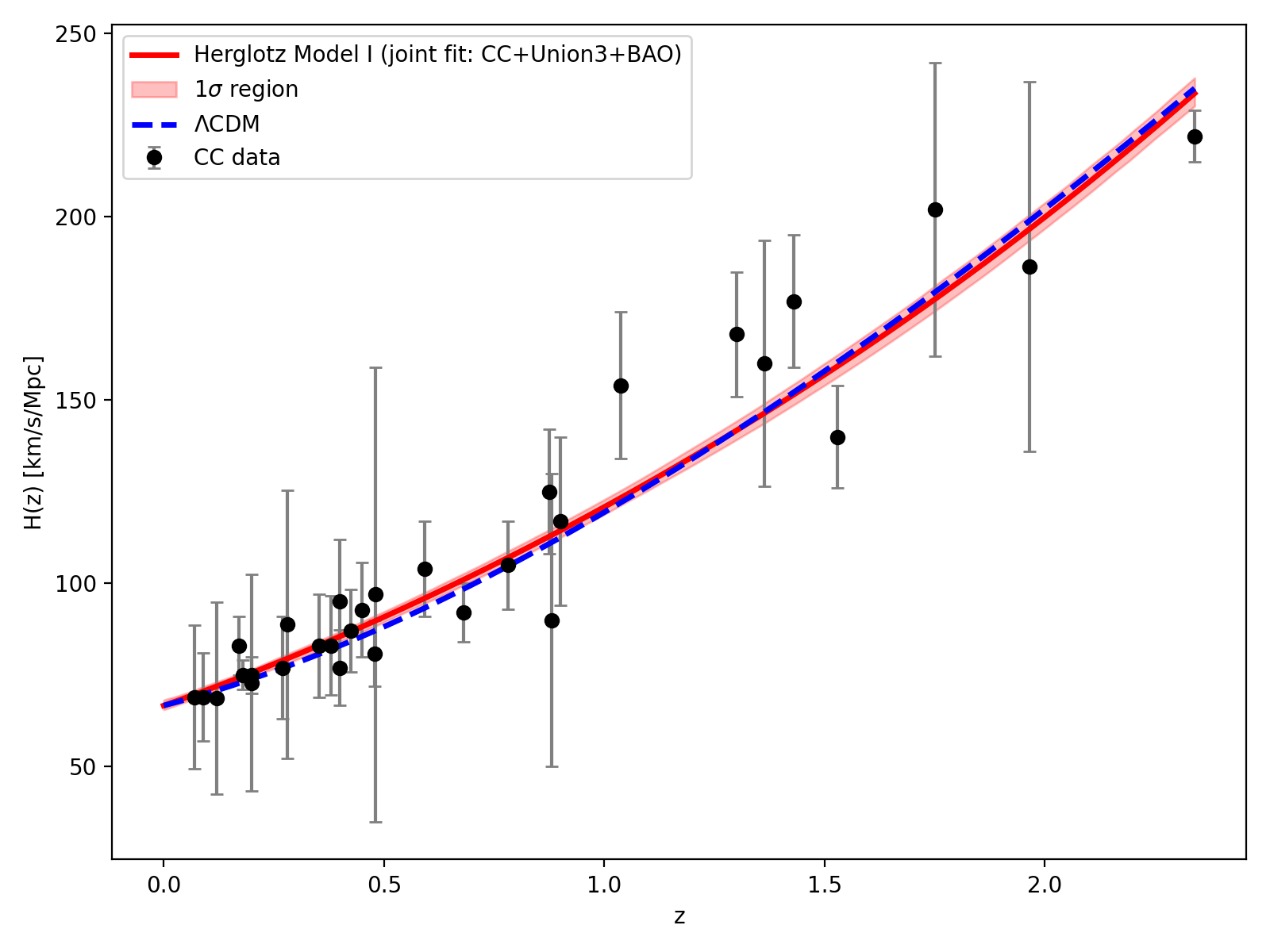}
\caption{Reconstructed Hubble expansion history for the combined
CC+Union3+DESI DR2 BAO analysis of the Herglotz Model. The solid
curve denotes the reconstruction using the posterior-median
parameters, the shaded region represents the $1\sigma$ posterior
uncertainty, and the dashed curve shows the spatially flat
$\Lambda$CDM prediction. The black points with error bars correspond
to the cosmic chronometer measurements.}
    \label{Combined_Hz.png}
\end{figure}

The corresponding reconstructed Hubble expansion history is shown in
Fig.~\ref{Combined_Hz.png}, where the best-fit Herglotz Model I is
compared with the cosmic-chronometer measurements, the $1\sigma$
posterior-predictive region, and the spatially flat $\Lambda$CDM
prediction evaluated using the same best-fit value of $H_0$. The
reconstructed Herglotz expansion history follows the $\Lambda$CDM
prediction closely over the redshift interval covered by the CC data,
$0\lesssim z\lesssim2.34$. A substantial fraction of the CC
measurements lies within or close to the $1\sigma$ posterior-predictive
region, with no pronounced systematic trend in the residuals over the
redshift range considered. This indicates that the joint best-fit
Herglotz Model provides an adequate description of the observed
expansion-rate measurements.

The close similarity between the reconstructed $H(z)$ and the
$\Lambda$CDM prediction is noteworthy when compared with the behaviour
of the derived cosmological diagnostics. In particular, the
deceleration parameter, effective equation-of-state parameter,
$Om(z)$ diagnostic, and statefinder parameters can exhibit more
pronounced deviations from the concordance model. This occurs because
these quantities depend on derivatives of the reconstructed expansion
history and therefore can reveal differences that are not readily
apparent in a direct comparison of $H(z)$ alone.

It is therefore important to emphasize that the close agreement of
the background Hubble expansion does not necessarily imply identical
cosmological dynamics. The Herglotz Model is governed by a coupled
system involving the dimensionless Hubble parameter $h(z)$ and the
Herglotz field $\Phi(z)$. Consequently, the reconstructed
higher-order cosmological quantities considered in the subsequent
sections provide complementary tests of the Herglotz modification
beyond the direct comparison of $H(z)$.

Taken together, the joint CC+Union3+DESI DR2 BAO analysis provides a
consistent observational reconstruction of the Herglotz Model. The
resulting background expansion history remains close to the
$\Lambda$CDM prediction over the redshift range probed by the CC
measurements, while the subsequent cosmological diagnostics allow
potential departures from the concordance dynamics to be examined in
greater detail.The posterior constraints obtained from the MCMC analysis are
summarized in Table~\ref{tab:parameter_constraints}. The table lists
the posterior median values and asymmetric $68\%$ credible intervals
for $(H_0,A,w,\Phi_0)$ for the CC, BAO, Union3, and combined datasets,
together with the adopted prior ranges. These constrained parameter
values are subsequently used to reconstruct the background expansion
history and evaluate the cosmological diagnostics of the Herglotz
Model.

\begin{table}[htbp]
\centering
\caption{Posterior constraints on the free parameters of the Herglotz
Model, $f(R,T)=R+\alpha T$, obtained from the individual and combined
observational datasets. The quoted values represent the posterior
medians with asymmetric $68\%$ credible intervals.}
\label{tab:parameter_constraints}
\renewcommand{\arraystretch}{1.5}
\resizebox{\textwidth}{!}{%
\begin{tabular}{lcccc}
\hline
Dataset &
$H_0$ [km s$^{-1}$ Mpc$^{-1}$] &
$A$ &
$w$ &
$\Phi_0$ \\
\hline

CC &
$64.1667^{+5.6198}_{-3.4654}$ &
$1.5066^{+2.1277}_{-1.0046}$ &
$-0.6243^{+0.2443}_{-0.4336}$ &
$0.2896^{+0.4558}_{-0.4818}$ \\[4pt]

BAO &
$66.5692^{+1.9208}_{-1.3913}$ &
$1.8121^{+1.2890}_{-0.8793}$ &
$-0.9269^{+0.2112}_{-0.2319}$ &
$-0.1410^{+0.3276}_{-0.2347}$ \\[4pt]

Union3 &
$72.0180^{+2.9965}_{-2.9240}$ &
$2.4111^{+1.4146}_{-1.3082}$ &
$-0.7464^{+0.1652}_{-0.2256}$ &
$0.0387^{+0.6117}_{-0.4192}$ \\[4pt]

Combined &
$66.6662^{+1.3177}_{-1.2838}$ &
$1.5665^{+0.6619}_{-0.5110}$ &
$-0.9924^{+0.1355}_{-0.1185}$ &
$-0.0620^{+0.1845}_{-0.1571}$ \\
\hline
\end{tabular}%
}

\vspace{2mm}
\begin{flushleft}
\footnotesize
Uniform prior ranges:
$50<H_0<85~{\rm km\,s^{-1}\,Mpc^{-1}}$,
$-0.9<A<5$,
$-3<w<1$, and
$-1<\Phi_0<1$.
\end{flushleft}
\end{table}

\section{Cosmological Diagnostics}
\label{subsec:Cosmological Diagnostics}

\subsection{Statefinder}

\begin{figure}[htbp]
\centering
\includegraphics[width=0.5199\linewidth]{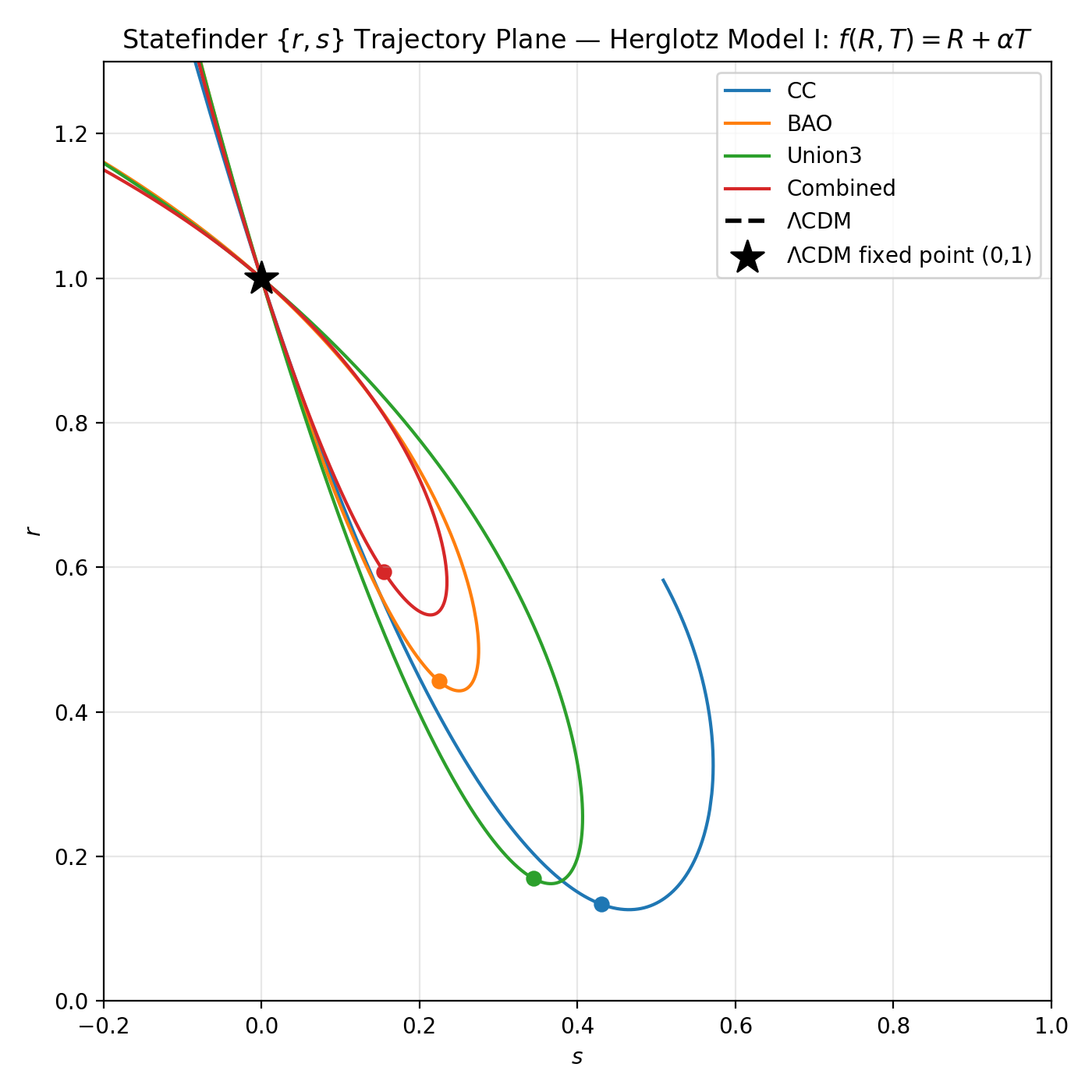}
\caption{Statefinder $\{r,s\}$ trajectories of the Herglotz-type $f(R,T)=R+\alpha T$ model for the CC, DESI DR2 BAO, Union3, and combined datasets. The black star denotes the $\Lambda$CDM fixed point $(s,r)=(0,1)$.}
    \label{statefinder.png}
\end{figure}

The statefinder $\{r,s\}$ diagnostic, originally introduced by \cite{sahni2003statefinder}
as a useful tool for distinguishing between different dark-energy scenarios beyond the Hubble and deceleration parameters, is presented in Fig.~\ref{statefinder.png} for the Model of Herglotz-type $f(R,T)=R+\alpha T$ gravity. The statefinder parameters are defined as
\begin{equation}
    r(z)=\frac{\dddot{a}}{aH^3},
\end{equation}
and
\begin{equation}
    s(z)=\frac{r(z)-1}
    {3\left(q(z)-\frac{1}{2}\right)}.
\end{equation}
For the spatially flat $\Lambda$CDM model, the statefinder
trajectory reduces to the fixed point
\begin{equation}
    (s,r)=(0,1),
\end{equation}
The four trajectories obtained from the CC, BAO, Union3, and Combined constraints exhibit distinct evolutionary paths and pass close to the $\Lambda$CDM fixed point during their evolution. As the redshift decreases toward the present epoch, the trajectories depart from this region and approach their respective $z=0$ locations, indicated by the filled circles. All four present-day points lie in the region $s>0$ and $r<1$, which is commonly associated with quintessence-like behaviour, rather than the $s<0$ and $r>1$ region typically associated with Chaplygin-gas-like evolution. The magnitude of the present-day deviation from $\Lambda$CDM is strongly dependent on the observational dataset. The CC-constrained reconstruction exhibits the largest departure from the concordance point, with $(s_0,r_0)\simeq(0.43,0.13)$, whereas the BAO-constrained trajectory remains comparatively closer to $\Lambda$CDM, with $(s_0,r_0)\simeq(0.23,0.44)$. The Union3 and Combined constraints yield intermediate behaviours, with $(s_0,r_0)\simeq(0.35,0.17)$ and $(s_0,r_0)\simeq(0.15,0.59)$, respectively. These results demonstrate that the statefinder behaviour of the Herglotz Model is sensitive to the observational dataset used to constrain its parameters. In particular, the displacement of all four present-day trajectories from the $\Lambda$CDM fixed point indicates that the corresponding best-fit reconstructions exhibit non-$\Lambda$CDM expansion dynamics at the present epoch.

\subsection{Effective Equation of State}

\begin{figure}[htbp]
\centering
\includegraphics[width=0.5199\linewidth]{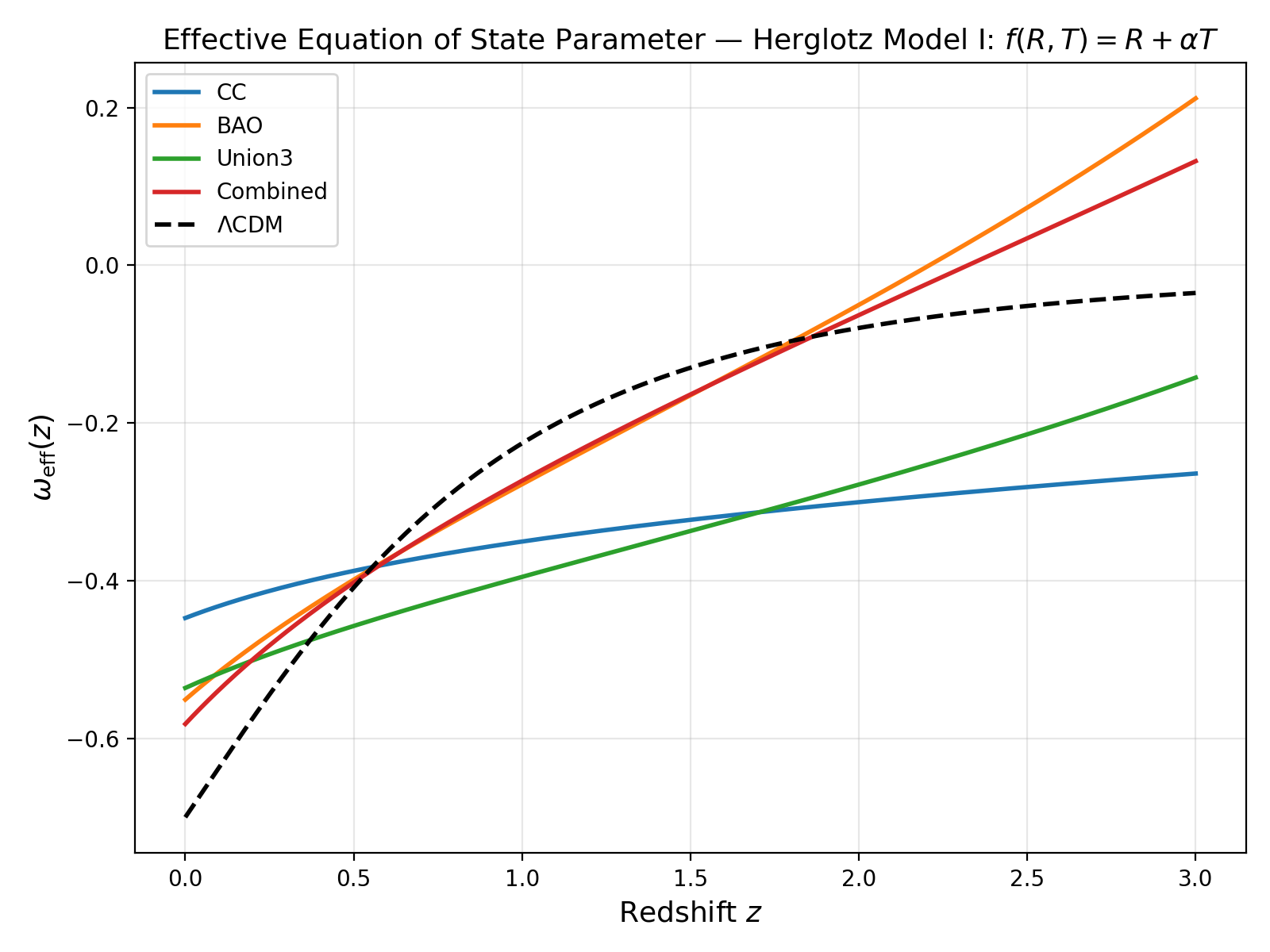}
\caption{Evolution of the effective equation-of-state parameter $\omega_{\rm eff}(z)$ for the Herglotz-type $f(R,T)=R+\alpha T$ model constrained by the CC, DESI DR2 BAO, Union3, and combined datasets. The dashed curve represents the $\Lambda$CDM prediction.}
    \label{w_eff_z.png}
\end{figure}

The evolution of the effective equation-of-state parameter,
$\omega_{\rm eff}(z)$, defined by
\begin{equation}
    \omega_{\rm eff}
    =\frac{p_{\rm eff}}{\rho_{\rm eff}}
    =\frac{2q-1}{3},
\end{equation}
and therefore directly determined by the reconstructed deceleration parameter, is presented in Fig.~\ref{w_eff_z.png} for the four best-fit realisations of Herglotz Model, together with the corresponding $\Lambda$CDM prediction. At the present epoch, all four reconstructions lie within the quintessence regime, $-1<\omega_{\rm eff}(0)<-1/3$\cite{Copeland:2006wr}. The Combined and BAO constraints give the most negative present-day values, $\omega_{\rm eff}(0)\simeq-0.58$ and $\simeq-0.56$, respectively, followed by the Union3 reconstruction with $\omega_{\rm eff}(0)\simeq-0.53$, while the CC-only constraint yields a comparatively larger value of $\omega_{\rm eff}(0)\simeq-0.44$. None of the reconstructed trajectories crosses the phantom divide, $\omega_{\rm eff}=-1$, over the redshift interval considered. Moreover, the Combined reconstruction remains comparatively closer to the $\Lambda$CDM behaviour at the present epoch, for which $\omega_{\rm eff}(0)\simeq-0.70$ for the adopted value of $\Omega_{m0}$.

As the redshift increases, the effective equation-of-state parameter generally evolves toward larger values, indicating a gradual departure from the negative-pressure regime at late times. The different observational constraints, however, lead to substantially different high-redshift behaviours. In particular, the BAO reconstruction crosses the matter-like value $\omega_{\rm eff}=0$ at approximately $z\simeq1.9$, while the Combined reconstruction crosses this value at a somewhat higher redshift. In contrast, the CC and Union3 reconstructions remain in the negative-$\omega_{\rm eff}$ regime throughout the plotted interval and reach approximately $\omega_{\rm eff}\simeq-0.27$ and $\simeq-0.14$, respectively, at $z=3$. The BAO and Combined trajectories subsequently enter the positive-$\omega_{\rm eff}$ region, reaching approximately $\omega_{\rm eff}\simeq0.21$ and $\simeq0.13$, respectively, at $z=3$. Although these values correspond to positive effective pressure, they should not be interpreted as a stiff-fluid regime, for which $\omega_{\rm eff}=1$.\\

The behaviour differs from that of $\Lambda$CDM, for which $\omega_{\rm eff}(z)$ evolves smoothly from a negative value at late times toward the matter-like limit $\omega_{\rm eff}\rightarrow0$ at high redshift. The comparatively rapid rise of the BAO and Combined reconstructions therefore indicates a stronger dataset-dependent modification of the effective cosmic expansion history at intermediate and high redshifts. The near coincidence of the BAO and Combined trajectories over a substantial portion of the redshift range suggests that the combined constraint is strongly influenced by the BAO information, whereas the comparatively flatter CC and Union3 trajectories indicate weaker constraints on the high-redshift evolution of the effective equation of state from these datasets individually. Overall, the results demonstrate that the inferred evolution of $\omega_{\rm eff}(z)$ in Herglotz Model is sensitive to the observational dataset, with all four reconstructions exhibiting quintessence-like behaviour at the present epoch but significantly different evolutionary trends toward higher redshift.

\subsection{Evolution of the Deceleration Parameter}

\begin{figure}[htbp]
\centering
\includegraphics[width=0.55\linewidth]{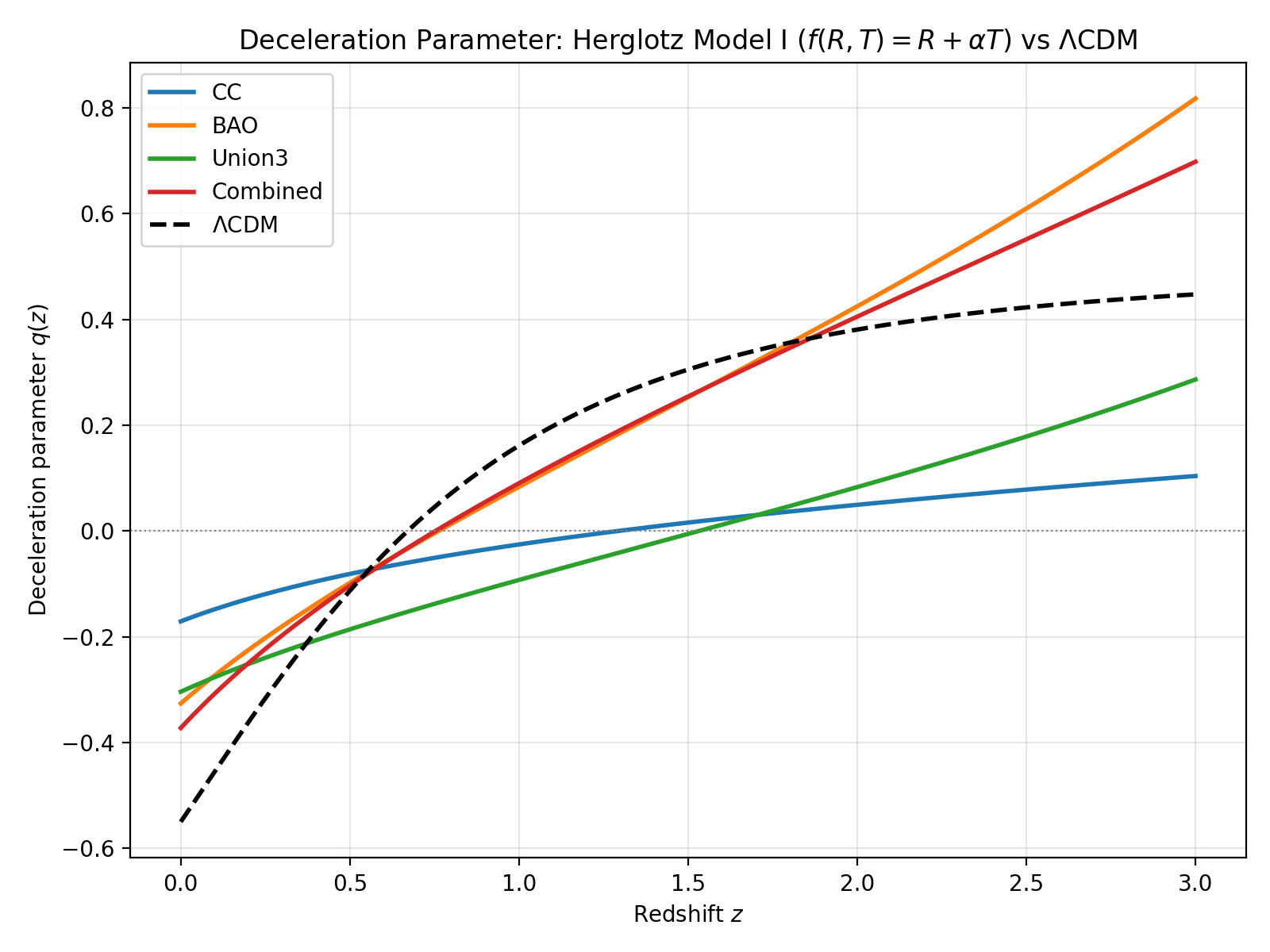}
\caption{Evolution of the deceleration parameter $q(z)$ for the Herglotz-type $f(R,T)=R+\alpha T$ model constrained by the CC, DESI DR2 BAO, Union3, and combined datasets. The dashed curve represents the $\Lambda$CDM prediction.}
    \label{q_z_comparison.png}
\end{figure}

The deceleration parameter,
\begin{equation}
    q(z)=-1+(1+z)\frac{h'(z)}{h(z)},
\end{equation}
reconstructed from the numerically integrated dimensionless Hubble parameter $h(z)$, is shown in Fig.\ref{q_z_comparison.png} for the four best-fit realisations of Herglotz Model together with the spatially flat $\Lambda$CDM prediction. At the present epoch, all four dataset combinations yield negative values of $q_0$, indicating an accelerating Universe at $z=0$. The magnitude of the present-day acceleration, however, varies appreciably among the different observational constraints. The CC-only reconstruction gives the weakest acceleration, with $q_0\simeq-0.17$, compared with $q_0\simeq-0.55$ for the adopted $\Lambda$CDM reference model. The BAO, Union3, and Combined constraints yield progressively more negative values, $q_0\simeq-0.32$, $-0.30$, and $-0.37$, respectively, bringing these reconstructions comparatively closer to the concordance prediction, although noticeable deviations remain.\\

As the redshift increases, all four Model trajectories evolve toward larger values of $q(z)$ and eventually cross the acceleration--deceleration boundary, $q(z)=0$. The corresponding transition redshift is strongly dependent on the observational dataset. The BAO and Combined reconstructions undergo the transition at approximately $z_t\simeq0.6$--$0.7$, close to the $\Lambda$CDM transition redshift of $z_t\simeq0.67$. In contrast, the CC and Union3 constraints predict substantially later transitions, at approximately $z_t\simeq1.3$ and $z_t\simeq1.7$, respectively. Thus, the inferred epoch of the transition from accelerated to decelerated expansion is significantly affected by the dataset used to constrain the model.\\

At higher redshifts, the differences among the reconstructed trajectories become increasingly pronounced. The BAO trajectory rises most rapidly, crossing above the $\Lambda$CDM prediction around $z\simeq1.8$--$2.0$ and reaching $q(z)\simeq0.82$ at $z=3$. The Combined reconstruction follows a qualitatively similar evolution and reaches $q(z)\simeq0.70$ at the upper end of the plotted range. By contrast, the CC and Union3 trajectories increase more gradually, reaching only $q(z)\simeq0.10$ and $q(z)\simeq0.29$, respectively, at $z=3$. Consequently, these two reconstructions remain below the $\Lambda$CDM prediction over much of the high-redshift interval considered, whereas the BAO and Combined constraints produce substantially stronger deceleration at higher redshifts. The separation between the BAO/Combined and CC/Union3 reconstructions is consistent with the dataset-dependent behaviour also observed in the $\omega_{\rm eff}(z)$ and $Om(z)$ diagnostics. Overall, Fig.~\ref{q_z_comparison.png} demonstrates that Herglotz Model predicts an accelerating present Universe for all four observational combinations, while the inferred transition redshift and subsequent high-redshift evolution of the deceleration parameter remain sensitive to the choice of dataset.

\subsection{Om(z) Diagnostic}

The $Om(z)$ diagnostic\cite{PhysRevD.78.103502},
\begin{equation}
    Om(z)=\frac{E^2(z)-1}{(1+z)^3-1},
    \qquad
    E(z)=\frac{H(z)}{H_0},
\end{equation}
provides a useful null test of the $\Lambda$CDM scenario based directly on the expansion history. For a spatially flat $\Lambda$CDM cosmology, $Om(z)$ remains constant and is equal to the present-day matter density parameter, $Om(z)=\Omega_{m0}$, irrespective of redshift. Consequently, any significant redshift dependence of $Om(z)$ indicates a departure from the $\Lambda$CDM expansion history. Moreover, the sign of the slope of the $Om(z)$ function is commonly used to characterize the effective dark-energy behaviour: a positive slope is associated with quintessence-like behaviour ($\omega_{\rm eff}>-1$), whereas a negative slope is associated with phantom-like behaviour ($\omega_{\rm eff}<-1$). It should be emphasized that this classification refers to the effective behaviour inferred from the $Om(z)$ diagnostic relative to the $\Lambda$CDM expansion history and should not be interpreted as a direct statement about the equation of state of the underlying dust component.\\

\begin{figure}[htbp]
\centering
\includegraphics[width=0.5199\linewidth]{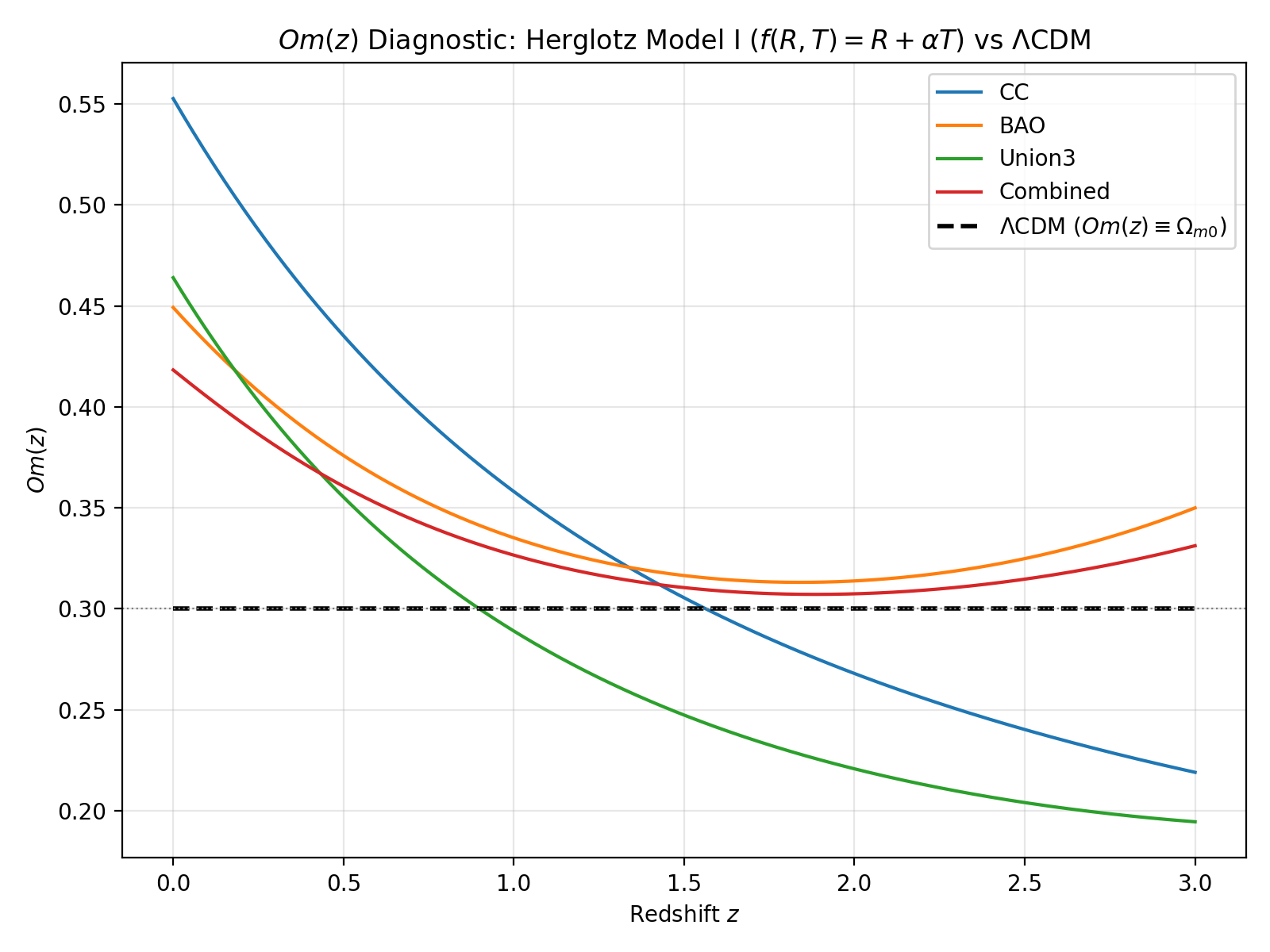}
\caption{Evolution of the $Om(z)$ diagnostic for the Herglotz-type $f(R,T)=R+\alpha T$ model constrained by the CC, DESI DR2 BAO, Union3, and combined datasets. The dashed line represents the $\Lambda$CDM value, $Om(z)=\Omega_{m0}=0.30$.}
    \label{Om_z_diagnostic.png}
\end{figure}

Figure~\ref{Om_z_diagnostic.png} presents the $Om(z)$ diagnostic for the four best-fit reconstructions of Herglotz Model, together with the spatially flat $\Lambda$CDM reference value $\Omega_{m0}=0.30$. The four reconstructions exhibit clear deviations from the $\Lambda$CDM reference over most of the redshift interval considered. At low redshift, the reconstructed $Om(z)$ values lie above the $\Lambda$CDM value, with the present-day values ranging approximately from $Om(0)\simeq0.42$ for the Combined constraint to $Om(0)\simeq0.55$ for the CC-only constraint. The curves subsequently decrease with increasing redshift over approximately $0\lesssim z\lesssim1.5$, corresponding to a negative $Om(z)$ slope and hence phantom-like effective behaviour according to the standard $Om$ diagnostic. The decrease is most pronounced for the CC reconstruction, which exhibits a reduction of approximately $0.3$ in $Om(z)$ by $z\simeq1.5$, whereas the BAO and Combined reconstructions flatten more rapidly.\\

At intermediate and higher redshifts, the dataset dependence becomes more pronounced. The BAO and Combined reconstructions develop shallow minima around $z\simeq1.5$--$2$, with $Om(z)$ approaching approximately $0.31$, close to the $\Lambda$CDM reference value. Beyond these minima, both trajectories turn upward, crossing the $\Lambda$CDM reference line and reaching approximately $Om(z)\simeq0.35$ and $Om(z)\simeq0.33$, respectively, at $z=3$. The corresponding change in the sign of the $Om(z)$ slope indicates a transition from phantom-like to quintessence-like effective behaviour within the redshift range considered. In contrast, the CC and Union3 reconstructions continue to decrease beyond $z\simeq2$, reaching approximately $Om(z)\simeq0.22$ and $Om(z)\simeq0.19$, respectively, at $z=3$, without exhibiting a clear turnover over the plotted interval.\\

The substantial separation of the reconstructed trajectories from the constant $\Lambda$CDM reference, together with their markedly different high-redshift behaviours, demonstrates the sensitivity of the $Om(z)$ diagnostic to the observational dataset used to constrain the Model. The CC and Union3 constraints favour a persistently decreasing $Om(z)$ over the redshift range considered, whereas the BAO and Combined constraints allow a turnover followed by an increasing $Om(z)$ at higher redshift. These differences, when considered together with the statefinder and effective equation-of-state analyses, indicate that the inferred departure of Herglotz Model from the concordance expansion history remains dependent on the observational combination employed.

\section{Conclusion}
\label{subsec:Conclusion}

In this work, we have investigated the late-time cosmological behaviour of Herglotz-type $f(R,T)$ gravity for the linear Model,
\begin{equation}
    f(R,T)=R+\alpha T,
\end{equation}
with particular emphasis on its observational viability and the dependence of the reconstructed cosmological evolution on the choice of observational data. Unlike conventional $f(R,T)$ models, the Herglotz-type formulation introduces an additional Herglotz field contribution to the cosmological background dynamics, resulting in a coupled system for the dimensionless Hubble parameter $h(z)$ and the Herglotz field $\Phi(z)$. Since the resulting coupled system does not yield a convenient closed-form expression for the Hubble parameter $H(z)$ under the effective equation-of-state closure adopted here, we solve the background equations numerically and incorporate the resulting numerical expansion history directly into the likelihood analysis.\\

We performed Bayesian parameter estimation for the four-dimensional parameter space $(H_0,A,w,\Phi_0)$, where $H_0$ denotes the present-day Hubble constant, $A$ characterises the coupling between the matter trace and the gravitational sector, $w$ represents the effective equation-of-state parameter entering the closure relation, and $\Phi_0$ is the present-day value of the Herglotz field. The analysis was carried out independently using Cosmic Chronometer (CC), DESI DR2 BAO, and Union3 supernova data, followed by a combined analysis of all three datasets. The individual observational probes lead to noticeably different posterior distributions, demonstrating that the inferred model parameters depend appreciably on the type of cosmological information employed. In particular, the CC analysis provides relatively weak constraints on $A$, $w$, and $\Phi_0$, while the BAO data provide stronger localization of the background parameters. The Union3 analysis favours a comparatively higher value of $H_0$, illustrating the dataset dependence of the inferred expansion scale. Combining the three complementary probes significantly reduces several of the parameter degeneracies and yields the representative constraint
\begin{equation}
    (H_0,A,w,\Phi_0)
    \simeq
    (66.7,\,1.57,\,-0.99,\,-0.06),
\end{equation}
for the joint reconstruction.\\

The reconstructed expansion history was further examined through several cosmological diagnostics. The deceleration parameter $q(z)$ indicates an accelerating Universe at the present epoch for all observational combinations considered, with the transition from acceleration to deceleration occurring at a dataset-dependent redshift. The BAO and Combined constraints predict transition redshifts closer to that of the $\Lambda$CDM model, whereas the CC and Union3 reconstructions exhibit comparatively later transitions. The effective equation-of-state parameter $\omega_{\rm eff}(z)$ likewise indicates quintessence-like behaviour at the present epoch for all four reconstructions, while their evolution toward higher redshift differs substantially depending on the dataset. None of the reconstructed trajectories crosses the phantom divide over the redshift range investigated.\\

The $Om(z)$ diagnostic provides further evidence for departures from the constant $\Lambda$CDM behaviour. While the different reconstructions exhibit qualitatively distinct redshift evolution, the BAO and Combined constraints develop a turnover at intermediate redshifts followed by an increasing $Om(z)$, whereas the CC and Union3 reconstructions remain predominantly decreasing over the considered interval. These differences highlight the sensitivity of the reconstructed effective expansion history to the observational dataset. Similarly, the statefinder $\{r,s\}$ analysis shows that the present-day reconstructions occupy locations away from the $\Lambda$CDM fixed point $(s,r)=(0,1)$, with the magnitude of the departure varying significantly among the different datasets. Thus, although the background $H(z)$ curves can remain close to the $\Lambda$CDM prediction, higher-order cosmographic diagnostics can reveal appreciable differences in the underlying expansion dynamics.\\

The combined $H(z)$ reconstruction provides a particularly interesting result. The best-fit Herglotz Model follows the $\Lambda$CDM expansion history closely over the redshift range covered by the available CC measurements, while the derived quantities $q(z)$, $\omega_{\rm eff}(z)$, $Om(z)$, and $\{r,s\}$ retain noticeable differences from the concordance model. This demonstrates that a direct comparison of $H(z)$ alone may not be sufficient to distinguish the Herglotz-type model from $\Lambda$CDM, whereas derivative-based and diagnostic quantities provide additional information about the underlying cosmological dynamics.\\

Overall, our results demonstrate that Herglotz-type $f(R,T)$ gravity with $f(R,T)=R+\alpha T$ provides an observationally acceptable description of the late-time background expansion, while its detailed cosmological behaviour remains sensitive to the observational data used for parameter estimation. The present analysis extends the original numerical viability study by replacing manually selected parameter values with a systematic Bayesian constraint of the model parameters using multiple complementary datasets. The resulting dataset dependence and the residual deviations from the $\Lambda$CDM statefinder and effective-fluid behaviour motivate further investigation using additional observational probes and model-selection statistics. In particular, future work incorporating larger and more homogeneous BAO and supernova compilations, gravitational-wave standard sirens, growth-of-structure measurements, and information criteria such as AIC and BIC would provide a more stringent assessment of whether Herglotz-type $f(R,T)$ gravity can be observationally distinguished from the standard $\Lambda$CDM cosmology.

\bibliographystyle{unsrt}
\bibliography{reference}
\end{document}